\documentclass[journal=nalefd,manuscript=letter,layout=traditional]{achemso}
\usepackage[version=3]{mhchem} 

\usepackage{bm}
\usepackage{graphicx}
\usepackage{caption}
\usepackage{booktabs}

\usepackage{pdflscape}
\usepackage{subcaption}
\usepackage{regexpatch}
\usepackage{hyperref}
\usepackage[T1]{fontenc}
\usepackage[utf8]{inputenc}
\usepackage{amsmath}
\usepackage{physics}
\usepackage{xargs}
\usepackage{longtable}
\usepackage[table,xcdraw]{xcolor}
\usepackage{colortbl}
\usepackage{multirow}
\usepackage{amsthm}%
\usepackage{mathrsfs}%

\mciteErrorOnUnknownfalse 
\hypersetup{pdfborder=0 0 0,colorlinks=true,citecolor=blue,linkcolor=blue}

\title{Spin-Polarized Electronic Structure and Chemical Bonding Data for 2,500+ Halide Double Perovskites}

\author{{Luc} {Walterbos}}
\affiliation{Department Materials Chemistry, Federal Institute of Materials Research and Testing, 12205 Berlin, Germany}
\alsoaffiliation{Institute of Condensed Matter Theory and Optics, Friedrich Schiller University, 07743 Jena, Germany}

\author{{Alex} {McEwan}}
\affiliation{MESA+ Institute for Nanotechnology, University of Twente, 7522 NB Enschede, The Netherlands}

\author{{Ravindra} {Shinde}}
\affiliation{MESA+ Institute for Nanotechnology, University of Twente, 7522 NB Enschede, The Netherlands}

\author{{Janine} {George}}
\affiliation{Department Materials Chemistry, Federal Institute of Materials Research and Testing, 12205 Berlin, Germany}
\alsoaffiliation{Institute of Condensed Matter Theory and Optics, Friedrich Schiller University, 07743 Jena, Germany}

\author{{Linn} {Leppert}}
\affiliation{School of Metallurgy and Materials, University of Birmingham, B15 2SE Birmingham, United Kingdom}
\alsoaffiliation{MESA+ Institute for Nanotechnology, University of Twente, 7522 NB Enschede, The Netherlands}
\email{l.leppert@bham.ac.uk}

\begin{document}
\maketitle

\begin{abstract}
Halide double perovskites (A$_2$BB'X$_6$) are a long-known class of materials that has recently been rediscovered for diverse applications, including photovoltaics, photocatalysis, and radiation detection. Their doubled unit cell provides immense chemical tunability, allowing the incorporation of magnetic ions and enabling access to a wide range of electronic-structure features, including different band-edge characters, alignments, and symmetries. Magnetic elements may further introduce spin degrees of freedom and magnetic behaviour, thereby broadening the functional landscape of these compounds.
Here, we present the first comprehensive database of spin-polarised electronic-structure data for all halide double perovskites predicted to be stable by the recently introduced $\tau$ tolerance factor by Bartel \textit{et al}. 
The dataset focuses on the Cs$_2$BB'X$_6$ family, with X = I, Br, Cl, and F, and includes density of states (DOS) for $>$2,500 compounds, calculated using hybrid-functional density functional theory. Among these, 719 compounds exhibit band gaps in the visible range and 118 display half-metallic character. In addition, we provide chemical-bonding analysis using \textsc{lobster}, which provides insights into orbital interactions across the dataset. To facilitate exploration, we further offer UMAP-based visualisations and an interactive app for systematic investigation of chemical composition, electronic structure, and magnetic properties.
\end{abstract}

\section{Background \& Summary}
Halide perovskites, with the chemical formula ABX$_3$ (where A and B are mono- and divalent cations, and X is a halide), crystallize in a network of corner-sharing BX$_6$ octahedra. Many halide perovskites show great potential for energy and quantum applications due to their tunable properties, facile synthesis and pronounced light-matter interactions. For example, the organic-inorganic perovskite methylammonium lead iodide ($\mathrm{MAPbI_3}$) \cite{Kojima2009} has favorable optoelectronic properties and is widely used in single and tandem solar cells with record power conversion efficiencies \cite{nrel2017}. Although most research in photovoltaics has focused on MAPbI$_3$ and related mixed-site compositions \cite{singh2017, Eperon2014a, brauer_comparing_2020, bi2018}, the halide perovskite family encompasses a vast number of materials, including halide double perovskites \cite{Wolf2021}, vacancy ordered perovskites \cite{Maughan2016}, dimensionally reduced perovskites \cite{Mitzi1995, Stoumpos2016a, Pedesseau2016, connorLayeredHalideDouble2018, dyksikBroadTunabilityCarrier2020}, and other perovskite-like structures with edge- and/or face-sharing octahedral connectivity \cite{Umeyama2020a, kammingaRoleConnectivityElectronic2017}. This enormous structural and compositional flexibility opens a broad material design space, allowing access to bespoke properties for a wide range of applications.

In this space, halide double perovskites (HDPs, also known as elpasolites) are a particularly interesting subclass because they accommodate two distinct B-site cations - or, in vacancy-ordered compositions, a single B-site cation alternating with a vacancy  \cite{Wolf2021}. The doubling of the unit cell, leading to a chemical formula of A$_2$BB'X$_6$, greatly expands the number of possible perovskite compositions since the two B sites can host elements with oxidation states from +I to +IV \cite{Faber2016, Filip2018}. HDPs were first discovered in the late 19th century, and many stable compositions have since been identified and synthesized. They typically crystallize in a cubic structure (space group $Fm\bar{3}m$). With the advent of halide perovskites for photovoltaic applications, renewed interest in HDPs has led to the discovery of many new compositions, for example, Cs$_2$AgBiBr$_6$, which is highly stable and has favorable optoelectronic properties \cite{Slavney2016a, Volonakis2016, McClure2016}. Beyond photovoltaics, the compositional freedom of HDPs allows broad tuning of structural, (opto-)electronic, and magnetic properties \cite{Volonakis2017, Luo2018, slavneySmallBandGapHalideDouble2018, xueChemicalControlMagnetic2022, jobsisConductionBandTuning2024}. This has generated interest in HDPs for applications ranging from photocatalysis \cite{muscarellaHalideDoublePerovskiteSemiconductors2022} to humidity sensors \cite{weng2019} and X-ray detectors \cite{panCs2AgBiBr6SinglecrystalXray2017}. Furthermore, some HDPs feature robust excitonic properties due to chemical confinement effects \cite{Palummo2020, Biega2021a, cuccoFineStructureExcitons2023, kavanaghFrenkelExcitonsVacancyOrdered2022, biegaChemicalMappingExcitons2023}, suggesting potential for quantum information and sensing applications.

Incorporating transition- and rare-earth elements at the B sites opens a largely unexplored dimension in HDP research, introducing magnetic degrees of freedom alongside structural and electronic ones \cite{klarbringElectronicStructureMagnetic2023, jobsisConductionBandTuning2024, singhExploringMagnetismLeadfree2023}. Although rare-earth HDPs have been studied since the 1960s, only a few compositions are known \cite{Wolf2021} and their electronic properties remain poorly understood. With the spin magnetic moment as an additional parameter, the tunability of HDPs expands, for example, allowing for spin-engineering of charge-carrier lifetimes and spintronic applications. Recent work has explored the incorporation of 3d transition metals in HDPs, leading to the discovery of Cs$_2$AgFeCl$_6$, which is antiferromagnetic at very low temperatures \cite{ningMagnetizingLeadfreeHalide2020}. Furthermore, first-principles studies using density functional theory (DFT) by Klarbring \textit{et al.} and Singh \textit{et al.} have focused on magnetism in HDPs, reporting a number of new compositions but restricting their high-throughput searches to transition metals from the 3d, 4d, and 5d series \cite{klarbringElectronicStructureMagnetic2023, singhExploringMagnetismLeadfree2023}. Other computational studies on HDPs included a broader range of compositions but did not consider spin-polarized electronic structures or magnetic properties \cite{Faber2016, bartelNewToleranceFactor2019, Filip2018}.

To bring this material space within reach of systematic exploration, we present the first mapping of spin states and spin-polarized electronic structures across the predicted HDP landscape. To achieve this, we have implemented an automated workflow that uses DFT with hybrid-functional accuracy and generated a database of spin-polarized electronic-structure data, including an in-depth bonding analysis. One of us has recently released a quantum chemical bonding database for inorganic materials\cite{naik_quantum-chemical_2023}, and used an extended database to demonstrate that bonding descriptors posses strong predictive power for phonon-related and elastic properties within machine learning frameworks\cite{naik_critical_2026}. Bonding analysis has also been show to be predictive for optoelectronic properties\cite{saleh_predictive_2025}, and has previously been applied in the context of Halide (Double) Perovskites. \cite{walsh_principles_2015,goesten_mirrors_2018,maughan_perspectives_2019}. Together, these results highlight the benefit of chemical bonding data in data-driven materials design.

Our database contains $>$2500 compositions whose properties can be explored via an interactive application. We use the dimensionality reduction technique UMAP to visualize composition–property correlations in this dataset and show that these correlations are encoded in the spin-polarized projected density of states of the materials. With this, our database provides an overview of electronic properties, including spin polarization and magnetic moments, across stable HDP compositions and can be used for materials discovery, the design of new functional HDPs with tailored properties, and as an input for machine-learning studies.

\section{Methods}

\begin{figure}
    \centering
    \includegraphics[width=0.9\linewidth]{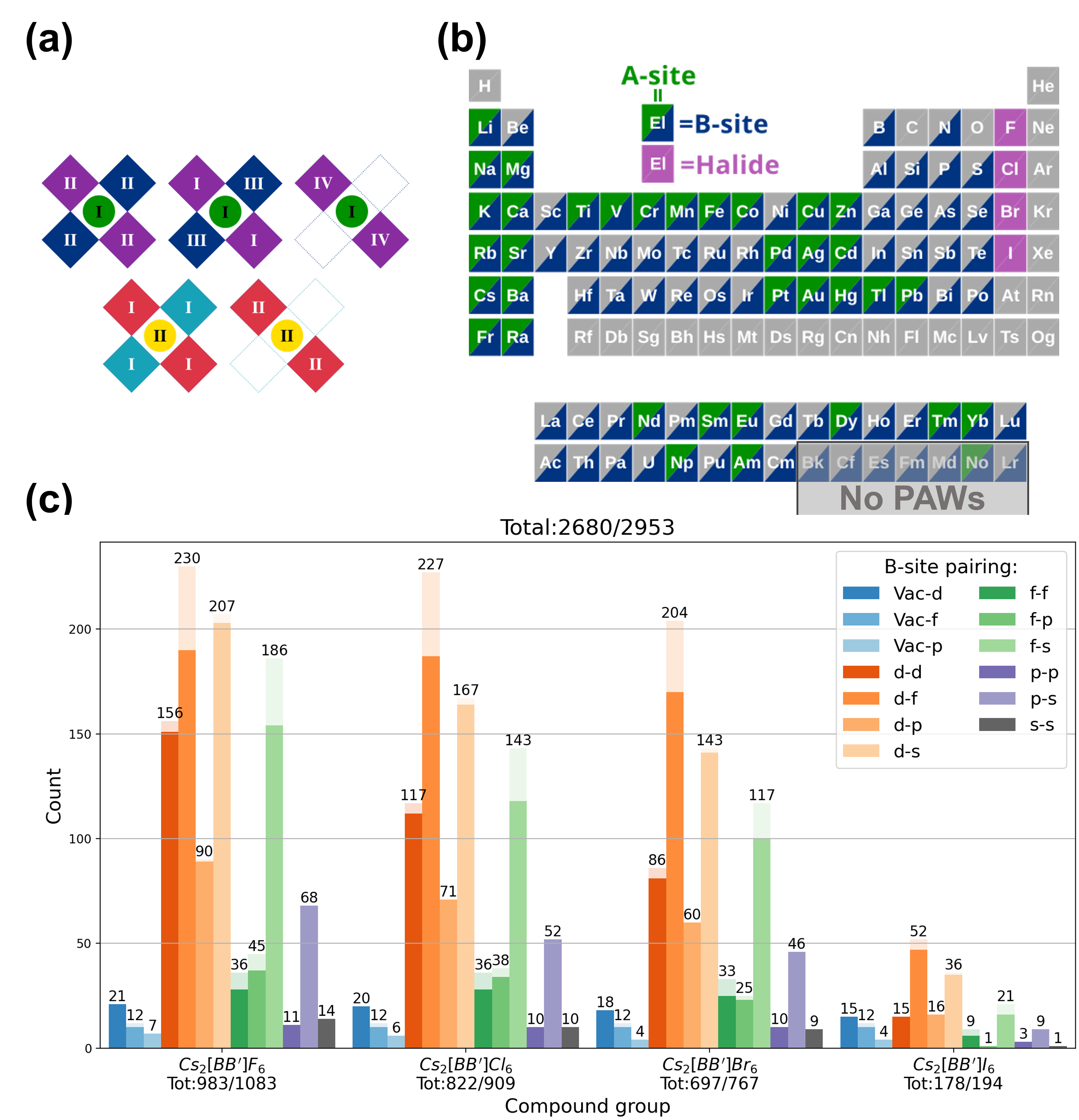}
    \caption{(a) Schematic representation of cationic oxidation state combinations that result in charge-neutral A$_2$BB'X$_6$, where X = I$^-$, Br$^-$, Cl$^-$, F$^-$. Halides are not shown. (b) Periodic table indicating which elements are featured in at least one predicted stable composition as A-site (with green upper left corner) or as B-site (by blue lower right corner). The X-sites are indicated with pink boxes.
    (c) For the compounds studied here (i.e., those with A=Cs$^+$), the bar plot shows the number of HDPs for each combination of blocks of the periodic table that the two B-sites originate from (Vac = vacancy-ordered). Due to the lack of available PAWs in \textsc{vasp} for elements from berkelium onward, a discrepancy arises between predicted (lower‑opacity bars, larger total number) and simulated (higher‑opacity bars, smaller total number) HDP compositions.}
    \label{fig:compselection}
\end{figure}

\subsection{Compound Selection}
To map the full chemical landscape of HDPs, we first enumerate all charge-neutral HDP compositions. Including vacancy-ordered double perovskites, the $6\times(-1)$ charge of the halide anions admits five possible cation charge combinations, illustrated in Figure~\ref{fig:compselection}a. Combining all elements with viable oxidation states under these constraints, yields 47,013 candidate compositions.

We then use a tolerance-factor-based assessment as stability criterion. For perovskites, several tolerance-factor definitions exist \cite{Kieslich2015, travisApplicationToleranceFactor2016}, based on the original definition by Goldschmidt \cite{Goldschmidt1926}. The Goldschmidt tolerance factor $t$ depends only on the ionic radii of the different species. The predictive power of $t$ is, however, severely limited for the heavier halides \cite{bartelNewToleranceFactor2019}. To overcome this limitation, Bartel \textit{et al.} introduced a new tolerance factor $\tau$, using SISSO (sure independence screening and sparsifying operator) \cite{ouyang_sisso_2018} on a database of experimentally realized perovskites \cite{bartelNewToleranceFactor2019}:
\begin{equation}
    \tau = \frac{r_X}{r_B} - n_A\left( n_A - \frac{r_A/r_B}{\ln{r_A/r_B}}\right),
    \label{eq:barteltau}
\end{equation}
where $r_i$ is the ionic radius of the ion on site $i$ and $n_A$ is the oxidation state of the A-site cation. Shannon's revised atomic radii \cite{shannon_revised_1970} were used, which were accessed through the Mendeleev python package \cite{mentel_mendeleev_2014}. Lower $\tau$ values are correlated with higher decomposition enthalpies, and Ref.~\citenum{bartelNewToleranceFactor2019} found that $\tau\leq 4.18$ predicted perovskite stability with a 91\% accuracy across all considered double perovskite compositions, and 84\% accuracy when restricted to HDPs \cite{bartel2020}.

For double perovskites, Ref.~\citenum{bartelNewToleranceFactor2019} computes $r_B$ in Equation~\ref{eq:barteltau} as the arithmetic mean of the B and B' ionic radii. This introduces an ambiguity for vacancy-ordered compositions: Basing $r_B$ on the B ionic radius alone underestimates the number of stable vacancy-ordered compositions, while averaging $r_B$ with a vacancy radius of zero overestimates it. We therefore apply a separate $\tau$ cutoff to vacancy-ordered HDPs, set as $\tau=5.46$, which is the highest $\tau$ among vacancy-ordered HDPs in the experimental data of Ref.~\citenum{brik_modeling_2011}.

Besides vacancy-ordered HDPs, cryolites\cite{Wolf2021,shuai_structure_2022} are another HDP subclass that was not considered by Bartel \textit{et al.}. In cryolites the A-site cation also occupies one of the B-sites, resulting in chemical formula A$_3$BX$_6$. In our database cryolites are represented as Cs$_2$CsBX$_6$. 
Apart from the inclusion of vacancy-ordered HDPs with a separate $\tau$ cutoff value and cryolites, our procedure for selecting compositions is the same as in Ref.~\citenum{bartelNewToleranceFactor2019}. Running this procedure on the 47,013 potential compositions resulted in 11,823 predicted stable compositions. Figure~\ref{fig:compselection}b shows which elements are featured in at least one predicted stable HDP composition.
Since A-site choice has limited direct influence on (opto‑)electronic properties, we fix the A-site cation and vary B‑site and X-site composition. Among the available A‑site cations, cesium permits the largest set of charge-balanced compositions. We therefore restrict the database to Cs-based HDPs. Under this constraint, 2,953 HDP compositions are predicted to be stable. The lack of \textsc{vasp} PAW (Projector Augmented Wave) pseudopotentials for elements with $Z\geq97$ further reduces this number to 2,680 compositions. Figure~\ref{fig:compselection}c provides an overview of compositional diversity in the dataset by showing the B‑cation pairings for each halide of the HDP, grouped by the periodic‑table blocks from which the B cations originate.

\subsection{Computational Workflow}
\begin{figure}
    \centering
    \includegraphics[width=0.5\linewidth]{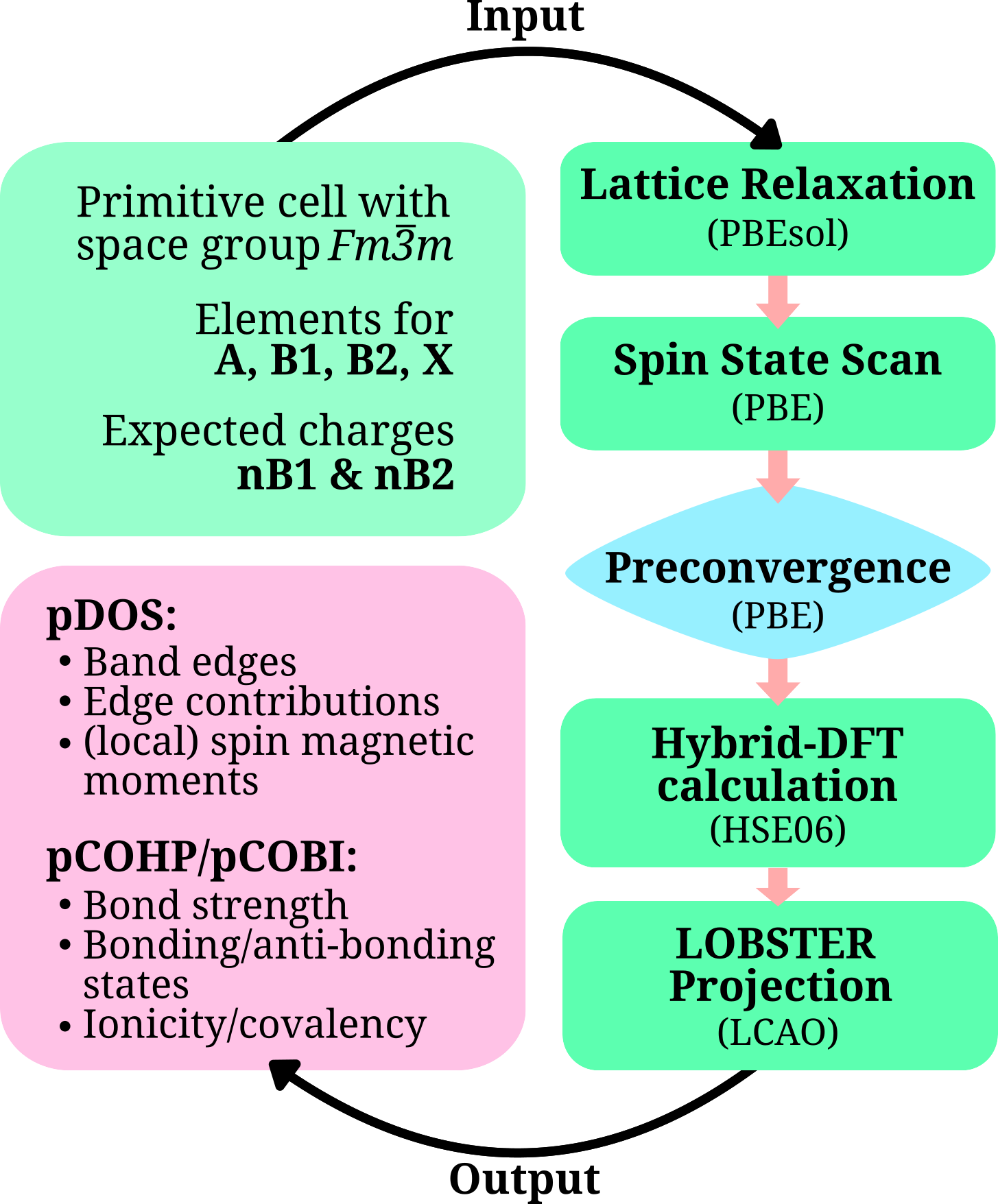}
    \caption{Overview of the automated workflow developed for this work.}
    \label{fig:WFoverview}
\end{figure}
Our automated workflow is schematically depicted in Figure~\ref{fig:WFoverview} and consists of four DFT steps using the \textsc{vasp} program package \cite{kresse_ab_1994,kresse_efficiency_1996,kresse_efficient_1996} with PAW pseudopotentials \cite{blochl_projector_1994,kresse_ultrasoft_1999}, followed by a projection onto atomic orbitals using \textsc{lobster} \cite{dronskowski_crystal_1993,deringer_crystal_2011,maintz_analytic_2013,maintz_lobster_2016,nelson_lobster_2020}. In general, the recommended PAW potentials included in \textsc{vasp 6.4} (potpaw.64) are used. For several lanthanides, the recommended PAW potentials treat the 4f electrons as frozen-core states. We refer to these as reduced-valence PAWs (RPAWs) and use them for structural relaxation to improve numerical stability. For subsequent electronic-structure calculations, where spin-related effects are of interest, we instead used PAW potentials that include 4f electrons as valence states. For cesium, calculations with the recommended potential (Cs\_sv) failed to converge. We therefore used the Cs\_sv\_GW potential, which provides a more accurate description of high-energy states via additional projectors. An overview of the PAW potentials used is provided in Table~\ref{tab:paw}.

All compositions were set up in the 10-atom primitive $Fm\Bar{3}m$ unit cell. The starting lattice parameter of $a=7.78$ {\AA} was determined from preliminary relaxation of 10 chloride compositions: We took the largest relaxed lattice parameter from this test set and expanded it by $\sim10\%$ to provide a conservative starting point. Fluoride, bromide, and iodide compositions were added to the workflow at a later stage; although bromide and iodide HDPs have larger equilibrium lattice parameters, using the same starting geometry did not lead to convergence issues.

For certain compositions, structural relaxation remained trapped in the high-symmetry configuration, with only minor ionic displacements before converging toward a local minimum and exhausting the maximum number of \textsc{vasp} ionic steps without achieving meaningful structural change. The workflow detects this outcome and restarts the relaxation with a larger initial lattice parameter ($a=8.06$ {\AA}). Other convergence failures were handled automatically by either tightening the electronic convergence criterion or cycling through alternative electronic minimization algorithms.

Structural relaxations were performed using PBEsol \cite{perdew_restoring_2008} with a force convergence threshold of 1\,meV/{\AA}. For these calculations, we used an increased plane-wave cutoff of $1.5$\texttt{ENMAX}, with a minimum of 500\,eV. For all subsequent DFT calculations, we used the largest default \texttt{ENMAX} value among the PAW potentials of the constituent elements, but not less than 350\,eV. A $4\times4\times4$ $\Gamma$-centered $k$-mesh was used throughout. 

In the second step of the workflow, we enumerated all combinations of high- and low-spin states (based on the $O_h$ point symmetry of the B-site) for transition metal, lanthanide, and actinide B-sites, including both parallel and antiparallel alignments of the magnetic sublattices. We did not consider non-collinear orderings, nor magnetic configurations requiring supercells beyond the 10-atom $Fm\bar{3}m$ primitive cell. Each configuration served as a starting point for a PBE \cite{perdew_generalized_1996} calculation, and the spin configuration with the lowest total energy was then selected for the remainder of the workflow. This was followed by a PBE pre-convergence run, and then the main electronic-structure calculation using the HSE06 \cite{heyd_hybrid_2003,heyd_erratum_2006} hybrid functional. The PBE pre-convergence and hybrid functional steps were performed using time-reversal symmetry only, to accommodate the projection in \textsc{lobster} \cite{nelson_lobster_2020}. The electronic convergence criterion was set to $10^{-4}$\,eV. The Brillouin zone was integrated using the tetrahedron method with Blöchl correction \cite{blochl_improved_1994}, and the number of energy points for the density of states (DOS) was set to 5000.

For the final step we used \textsc{lobster} to project the hybrid functional wavefunctions onto a linear combination of atomic orbitals (LCAO). We used the basis function set \textsc{pbeVaspFit2015} \cite{maintz_lobster_2016}. Using a minimal basis set, which only includes states occupied by valence electrons, is generally best practice, but several elements have additional polarization functions available. We assessed projection quality and the need for additional basis functions using the absolute charge spilling and band-overlap parameters. Our workflow started with projecting onto the minimal basis set. If the minimal basis yielded an absolute charge spilling above 3\% or the orthonormalization step resulted in a matrix with a maximum deviation from identity exceeding 0.1 (indicating overlapping bands), the workflow calculated additional basis sets until either both conditions were met or it ran out of additional functions. Several additional basis-set projections were performed under the band-overlap condition, but the ultimate choice of basis set was based on the absolute charge spilling criterion alone (see Technical Validation). The band-overlaps criterion was applied as a precautionary check to flag potentially non-physical projections. Corresponding additional projections are retained in the database.

A technical subtlety concerns the cesium basis set used in the \textsc{lobster} projections. We found that removing the Cs 6s basis functions substantially improved the projection quality for outlier systems (i.e., compositions with large absolute charge spilling, or low charge spilling but with localized charges summing to a value far from charge-neutrality). The origin of these deviations is likely related to the use of the Cs\_sv\_GW potential, whose projector structure differs from that assumed in the \textsc{pbeVaspFit2015} basis. Excluding the 6s basis functions solved all observed projection issues and produced equal or slightly improved results for compositions that were already well behaved. Given that the Cs $6s$ state is expected to be unoccupied and weakly involved in bonding in these materials, we adopted the reduced cesium basis as the standard.

The energy range for the DOS and other chemical bonding quantities was automatically matched to the HSE06 DOS range. To ensure an accurate representation of the projected DOS and chemical bonding indicators, we increased the number of energy points to 6000 in LOBSTER. We invoked the \textsc{cohpGenerator} keyword separately for each B-X pair, with a maximum distance of 4\,{\AA}, restricting the bonding analysis to nearest-neighbor B-X interactions. 

\subsection{Chemical Bonding Analysis}
\textsc{lobster} projects the PAW wavefunctions onto a Slater-type orbital local basis set. The projection onto LCAO improves the quality of several localized quantities, like atomic charges\cite{ertural_development_2019} and site-projected DOS (pDOS) \cite{maintz_analytic_2013}, compared to \textsc{vasp}'s standard projection method, which uses either a user-defined or PAW projector radius to define the projection sphere. \textsc{lobster} then determines, for each atomic orbital, the \textbf{k}-dependent expansion coefficients with respect to the plane-wave Bloch states, yielding the LCAO crystal orbital. 

Combined with the overlap integral between two atomic orbitals, $S_{\mu \nu} = \int \phi_\mu^*\phi_\nu \mathrm{d}\tau$, the coefficients lead to the first quantum-chemical bonding indicator: the projected crystal-orbital overlap population (pCOOP), giving an energy-resolved picture of where orbitals are overlapping, and which interactions are bonding/antibonding:
\begin{equation}
    \mathrm{pCOOP}_{\mu\nu}(E) = S_{\mu\nu}\sum_{j,\mathbf{k}}w_\mathbf{k} \operatorname{Re}(c^*_{\mu,j\mathbf{k}}c_{\nu,j\mathbf{k}})\cdot\delta(\varepsilon_j(\mathbf{k}) - E)
    \label{eq:pCOOP}
\end{equation}
Here $\mu,\nu$ indicate atomic orbitals located on different atoms, $\varepsilon_j$ is the energy eigenvalue of band $j$, $c_{\mu,j\mathbf k}$ are the coefficients of the LCAOs, and $w_\mathbf{k}$ is the $\mathbf k$-point weight. Integrating pCOOP($E$) up to the Fermi level yields the ICOOP, which corresponds to the number of electrons populating the bond. The pCOOP can be understood as an overlap-weighted density of states.

Another bonding indicator is the projected crystal-orbital Hamiltonian population (pCOHP), for which the LCAO basis is orthonormalized using Löwdin's symmetric orthonormalization (LSO) \cite{maintz_analytic_2013,lowdin_nonorthogonality_1950}, yielding the orthonormalized wavefunctions $\mathbf{\chi_k'}$ and coefficients $\mathbf{C_k'}$ respectively. The pCOHP is given by,
\begin{equation}
    \mathrm{pCOHP}_{\mu\nu}(E) = H_{\mu\nu}\sum_{j,\mathbf{k}}w_\mathbf{k} \operatorname{Re}(c'^*_{\mu,j\mathbf{k}}c'_{\nu,j\mathbf{k}})\cdot\delta(\varepsilon_j(\mathbf{k}) - E)
    \label{eq:pCOHP}
\end{equation}
where $H_{\mu\nu}$ is the Kohn-Sham Hamiltonian reconstructed in the orthonormalized LCAO basis, and $c'_{\mu,j\mathbf k}$ the LSO-LCAO coefficients. The pCOHP integrated up to the Fermi level (ICOHP) corresponds to the energy contained in the bond, i.e., the bond strength. The pCOHP can be understood as a Hamiltonian-weighted density of states that partitions the Kohn-Sham band energy.

Replacing the overlap matrix element $S_{\mu \nu}$ or Hamiltonian matrix element $H_{\mu \nu}$ with the density matrix element $P_{\mu\nu}$ gives the crystal-orbital bonding index (COBI)\cite{muller_crystal_2021}:
\begin{equation}
    \mathrm{COBI}_{\mu\nu}(E) = P_{\mu\nu}\sum_{j,\mathbf{k}}w_\mathbf{k} \operatorname{Re}(c^*_{\mu,j\mathbf{k}}c_{\nu,j\mathbf{k}})\cdot\delta(\varepsilon_j(\mathbf{k}) - E)
    \label{eq:COBI}
\end{equation}
The integrated value of the COBI, the ICOBI, corresponds to the classical bond order, with low values indicating a mostly ionic bond and values close to one (or two for double bonds) indicating a mostly covalent bond. 

Additionally, \textsc{lobster} computes orbital and gross populations, atomic charges, and Madelung energies based on both the Mulliken\cite{Mulliken1955} and Löwdin\cite{lowdin_nonorthogonality_1950} partitioning schemes. 

For this database, we omitted the pCOOP calculation (keyword skipCOOP) to reduce storage requirements, since bonding/antibonding information is also contained in the pCOHP and COBI. The initial LCAO basis can occasionally lead to negative occupations in orbital populations and the pDOS due to numerical projection errors. To ensure physically meaningful populations and consistent results across the dataset, the analysis of the pDOS and charges presented here is based on the LSO basis. The analysis scripts also supports pDOS analysis from \textsc{vasp} or the original \textsc{lobster} basis, as well as Mulliken-scheme population analysis.

\subsection{Generating Data Records}
As detailed above, we use the Bartel $\tau$-factor as our primary stability criterion. For comparison, the database also reports the generalized Goldschmidt ($t$) factor introduced by Filip and Giustino \cite{Filip2018}. Although the generalized $t$ has lower reported accuracy than $\tau$ for predicting perovskite stability, we find it informative for HDPs, because it captures both the average ionic radius of the two B-sites (through the average octahedral factor $\bar{\mu}=(r_B+r_{B'})/2r_X$) and the difference between them (through the octahedral mismatch $\Delta\mu=|r_B-r_{B'}|/2r_X$). Ref.~\citenum{Filip2018} combines $t$, $\bar{\mu}$, and $\Delta \mu$ into eight geometric inequalities, and translate violations of these inequalities into a stability prediction via volumetric integration over uncertainties in the ionic radii. We instead compute $t$, $\bar{\mu}$, and $\Delta \mu$ directly from the tabulated Shannon radii and report which (if any) of the geometric limits are violated, using the notation of Ref.~\citenum{Filip2018}.

For the relaxed structures, we report B-X bond lengths together with the X-X distances between opposing X-anions in each octahedron. The latter remains well-defined for vacancy-ordered HDPs (where one B-site is empty and the B-X distance is undefined) and thus enables direct structural comparison across standard and vacancy-ordered compositions.

The contributions near band edges were analyzed from the projected DOS in \textsc{lobster} (in the LSO basis). Since all A- and X-sites are symmetrically equivalent in our structural models, we summed over their respective pDOSs. We determined band-edge contributions within an analysis window of 0.5\,eV below the valence band maximum (VBM) and 0.5\,eV above the conduction band minimum (CBM). Within this window, all states were integrated, and for each species we computed its contribution to the integrated states and identified the dominant orbital.
The contributions to the VBM and CBM were determined separately for the spin-up and spin-down channels. We selected the highest VBM and lowest CBM as the global band edges across both spin channels, and stored the data in a separate 'combined' spin channel. Storing all three sets - spin-up, spin-down, and combined - allows us to identify spin-forbidden transitions, half-metallic compositions, and the band gap of the non-conducting spin channel in half-metals. Furthermore, we report the charges and the orbital gross population of each site to determine the site-specific spin magnetic moments. Owing to the simplified treatment of magnetism in our workflow, these values are not expected to be quantitatively accurate. They nonetheless serve as useful indicators when screening for compositions with interesting spin properties.

In addition to the standard quantities provided by the \textsc{lobster} analysis, we introduce a bonding descriptor specific to the HDP geometry. Our bonding analysis considers only nearest-neighbor B-X bonds. Therefore, we provide analysis only for the properties of these bonds, which are aligned with the Cartesian axes in the cubic structure imposed by our 10-atom primitive cell. This means that the cubic harmonics from \textsc{lobster} are properly aligned by default, enabling us to identify differences in bonding in the $x$, $y$, and $z$ directions. The projected COHP (pCOHB) and pCOBI (and their integrated values) are averaged over the two bonds along each direction and over all six B-X bonds. The B-X bonds along one axis are not always the same in the positive and negative directions. To quantify this, we use the asymmetry index introduced by Belli \textit{et al.}\cite{belli_chemical_2025} and implemented in LobsterPy \cite{naik_lobsterpy_2024} (which we dub axial-asymmetry index to distinguish it from the other asymmetry descriptor introduced for this work): 
\begin{equation}
    \mathrm{axial\_asym\_index} = |\mathbf S_i| = \left| \frac{1}{B_i} \sum_{a=1}^{B_i} \mathrm{ICO..}(i,a)\mathbf{e}_{i,a} \right| 
\end{equation}
where $B_i$ is the number of bonds from site $i$, $\mathbf e_{i,a}$ is the unit vector along bond number $a$ originating from site $i$, and ICO.. refers to any of ICOOP, ICOHP, or ICOBI. The axial-asymmetry index is the norm of the absolute differences in bond strength/index along each axis; when applied to the ICOBIs, it yields a maximum value of 0.0032 in our dataset. To quantify whether there is any asymmetry between the Cartesian axes, we introduce the directional-asymmetry index, given by:
\begin{equation}
    \mathrm{directional\_asym\_index} = \frac{\max(\mathrm{ICO.._{(x/y/z)}}) - \min(\mathrm{ICO.._{(x/y/z)}})}{\mathrm{ICO.._{avg}}}
\end{equation}
which gives the difference between the strongest and weakest bond relative to the six-bond average. Using the directional-asymmetry index on the ICOBIs gives a maximum value of 1.49 in our dataset.

\subsection{UMAP projection}
To visualize underlying trends and facilitate navigation of our database, we used the Uniform Manifold Approximation and Projection (UMAP) method \cite{mcinnes_umap_2020}, a non-linear dimensionality reduction algorithm that is similar to t-distributed Stochastic Neighbor Embedding (t-SNE) \cite{van_der_maaten_visualizing_2008} but with more favorable scaling. UMAP calculates the distance from a point $x_i$ to its $N$ nearest neighbors using a locally varying distance metric, producing a directed weighted graph. This graph is then symmetrized into an undirected graph. Next, a lower-dimensional (in our case, 2D) representation is initialized, and then the cross-entropy between the high- and low-dimensional representations is minimized. Two key UMAP parameters are the distance metric and the number of neighbors.

Since our aim is to visualize trends in the electronic-structure data directly, we used the pDOS and pCOHP data as input for UMAP. We applied a Gaussian smearing of 0.05\,eV to these data to mitigate disproportionate contributions of very sharp peaks in the pDOS and pCOHP to the distance measure. Additionally, we restricted the input to states within 5\,eV of the band edges, i.e., in the range [VBM - 5\,eV, CBM + 5\,eV], to exclude deep core state and high-energy conduction states that are not chemically informative. The VBM was shifted to 0\,eV across the dataset. To ensure the energy axis and intervals are equivalent for all compositions, we used the binning method introduced in Refs.~\citenum{purcell_accelerating_2023,kuban_madas_2024}. Overall, this gives an energy axis ranging from -5.00\,eV up to 14.83\,eV with 0.1\,eV intervals, resulting in a 199-dimensional vector for each pDOS/pCOHP curve selected for UMAP input. Binned pDOS and pCOHP curves were saved for each atomic site and spin channel to simplify testing different combinations of input data for the projection. 

To represent the compositions in the UMAP projection, several options exist based on the computed pDOS for each site and the B-X pCOHP curves. Additionally, different embeddings can be merged (using their union (+), intersection (*), and contrast (-)), combining the fuzzy edge weights between pairs of points across constituent embeddings. This allows us, for example, to compute separate embeddings from the B1 pDOS and B2 pDOS and then intersect them. Our UMAP projections are based on the Euclidean and Manhattan metrics, their standardized versions (weighted with the inverse standard deviation), the Bray-Curtis metric ($\sum_i|u_i-v_i|/\sum_i|u_i+v_i|$), and the cosine metric. We calculated each metric using the nearest-neighbor values $5,15,25,50,100$. As input to our projection, we tested the following combinations:
\begin{itemize}
    \item Total DOS (spin-up \& down concatenated) (TDOS)
    \item Intersection of B1-pDOS (spin-up \& down concatenated) and B2-pDOS (spin-up \& down concatenated)(B-pDOS)
    \item Intersection of B1 and B2 spin-up pDOS (concatenated) and B1 and B2 spin-down pDOS (concatenated) (alternative B-pDOS)
    \item Intersection of B1 spin-up, B1 spin-down, B2 spin-up, and B2 spin-down pDOS (separated B-pDOS)
    \item Intersection of B1-X average COHP with B2-X average COHP (B-X pCOHP)
\end{itemize}
and also a normalized version of each combination, for which the area under the binned curves was set to 1. 

The examples above were chosen to showcase how these data representation choices affect the projection, and are not intended to be final. Users of the database are encouraged to explore further options. For this purpose, we developed an interactive viewer using Dash (\url{www.dash.plotly.com}) that enables rapid switching between precomputed projections and coloring by computed material properties. Selecting a point in the embedding reveals the corresponding pDOS and pCOHP/pCOBI curves together with basic electronic-structure data for that specific material. This functionality helps explore different projection setups and evaluate whether the resulting embedding reflects meaningful chemical or physical trends. All precomputed projections are based on UMAP, but the interactive viewer can, in principle, be used with any dimensionality reduction algorithm, as long as the projection is two-dimensional and the projection is saved with the same composition indexing as the database and similar table header structure. 

\section{Data Records}
Of the 2,680 compositions fed into the workflow, 138 encountered convergence issues, leaving our database with Data Records for 2,542 HDP compositions spanning the chemical landscape.
The Data Records are separated into four categories: basic data on the input species and the \textsc{lobster} basis functions (Table~\ref{tab:basicinfo}); structural data, including lattice vectors and interatomic distances (Table~\ref{tab:structuralinfo}); band edge data, including band gaps and band-edge orbital character for spin-up, spin-down, and combined spin channels (Table~\ref{tab:bandedgeinfo}); and bonding descriptors (Table~\ref{tab:lobsterinfo}). The information from Tables~\ref{tab:basicinfo} - \ref{tab:lobsterinfo} is combined into a single \textsc{HDP\_CombinedInfo.csv} file for convenience, with only the 'combined' spin channel from Table~\ref{tab:bandedgeinfo} included. In this combined file, the site (with orbital character) with the largest contribution to the VBM and CBM edges is also extracted, and was used to create the coloring scheme in Figure~\ref{fig:bgmagmomplot}a.

All calculation data, including all \textsc{lobster} projections performed with additional basis functions, are made accessible through NOMAD\cite{scheidgen_nomad_2023} and the workflow was parsed with the NOMAD Utility Workflows plugin\cite{rudzinski_nomad_2026}. The dataset is publicly available via NOMAD \cite{luc_walterbos_nomad_nodate}. Data processing was implemented using several functionalities from PyMatGen\cite{ong_python_2013} and LobsterPy\cite{naik_lobsterpy_2024}. All scripts, the extracted data, and a list of the compositions that failed to converge are available on Zenodo \cite{lwalterbos_luccerboihdp_workflow_analysis_2026}. The interactive UMAP viewer, including the precomputed projections, is available on GitHub \cite{lwalterbos_luccerboihdp_umap_explorer_github} and is also archived on Zenodo\cite{lwalterbos_luccerboihdp_umap_explorer_2026}.

\section{Data Overview}
\begin{figure}
    \centering
    \includegraphics[width=0.9\linewidth]{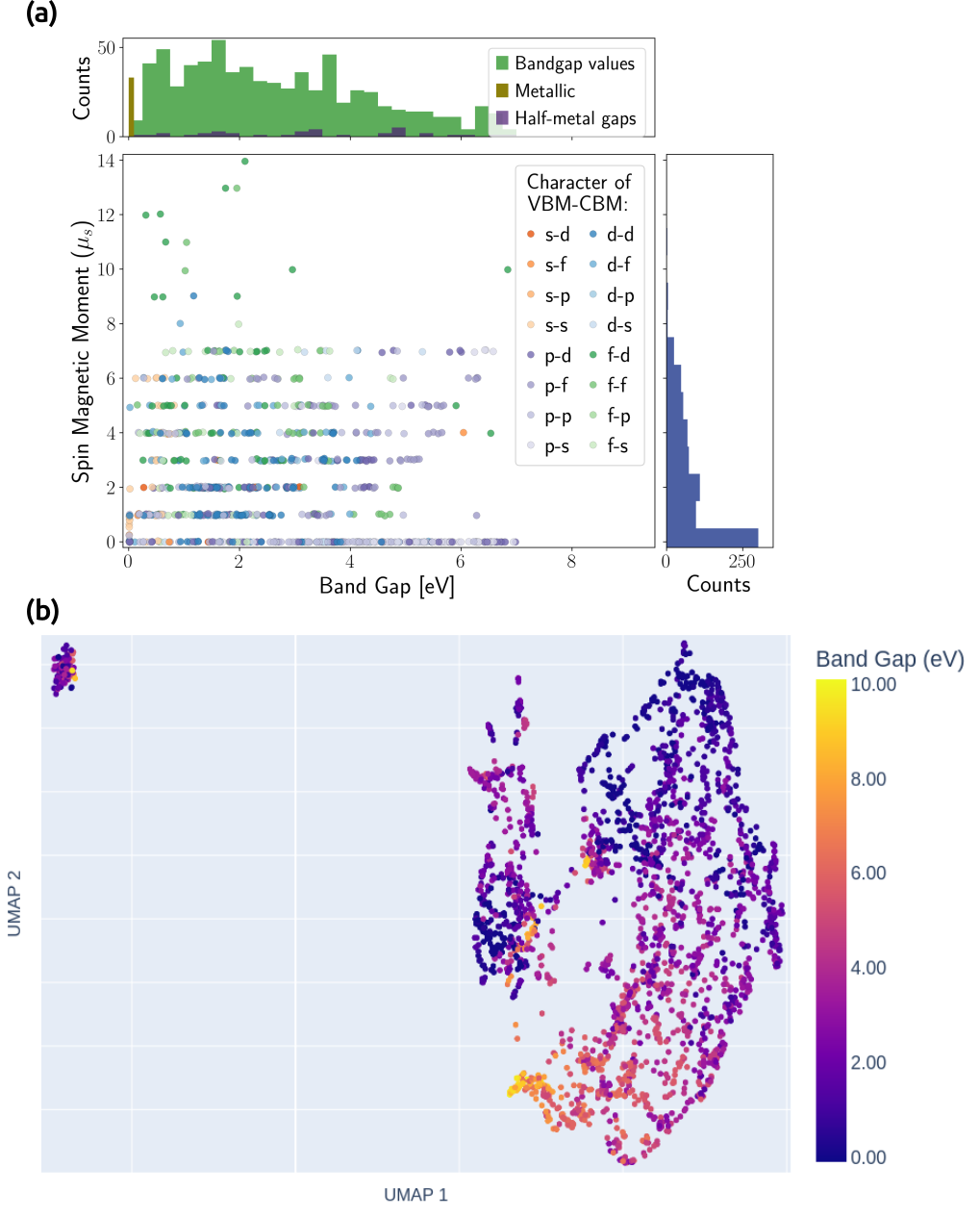}
    \caption{(a) Scatter plot of the net spin magnetic moment vs. band gap values for chloride HDPs. The color of the data points is determined by the orbital character of the valence band maximum (orange, purple, blue, green corresponding to s-, p-, d-, f-orbital character, respectively), whereas the color tone indicates the orbital character of the CBM. (b) UMAP projection based on the separated B-pDOS data using the Bray-Curtis metric with n\_neighbors=25, colored by the band gap values. }
    \label{fig:bgmagmomplot}
\end{figure}

Figure~\ref{fig:bgmagmomplot} provides representative views of the dataset. Figure~\ref{fig:bgmagmomplot}a shows a scatter plot of the net magnetic moments and band gaps for the Cs$_2$[BB']Cl$_6$ compositions, highlighting the diversity of electronic properties in this chloride-based subset of the data. Figure~\ref{fig:bgmagmomplot}b shows a selected UMAP projection of the separated pDOS, computed using the Bray-Curtis metric. Figure~\ref{fig:bgmagmomplot}b is colored by band gap values, revealing an underlying structure. The projection shows an isolated cluster in the top-left corner, comprising all vacancy-ordered HDP entries. Figure~\ref{fig:bgmagmomplot}b illustrates that the pDOS alone carries enough information for UMAP to reveal underlying structure in a metric of interest and to separate a distinct subgroup of HDPs. 

\begin{figure}
    \centering
    \includegraphics[width=0.9\linewidth]{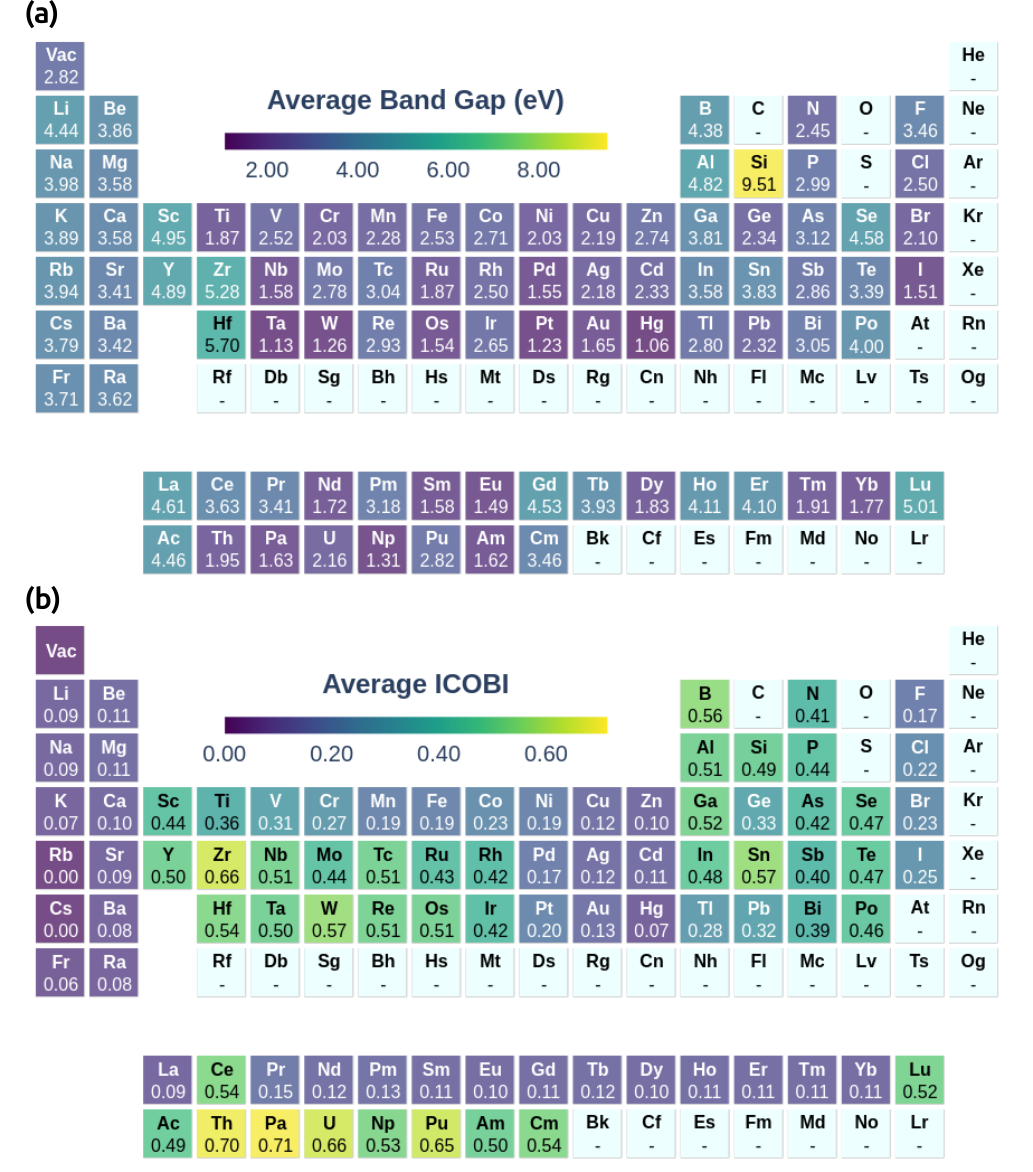}
    \caption{(a) Periodic table showing the average band gap for compositions associated with each element (in the case of Cs only compositions with Cs on a B-site are included). (b) Periodic table showing the average ICOBI value for each element. The figures were created using PyMatViz\cite{riebesell_pymatviz_2022} }
    \label{fig:ptables}
\end{figure}

Figures~\ref{fig:ptables}a and \ref{fig:ptables}d show the band gap and ICOBI values per element, averaged over the number of compositions in which they are incorporated on a B-site or X-site. As expected from electronegativity trends, fluorine-containing HDPs exhibit the lowest ICOBI values and the highest band gaps among the halides, reflecting their predominantly ionic character. Cl-, Br- and I-containing HDPs follow in sequence, with progressively higher ICOBI values and lower band gaps.

\section{Technical Validation}
\subsection{Completeness of Chemical Landscape}
Our goal was to cover the full chemical landscape of cesium HDPs, yet the database inevitably contains some gaps. Of the compositions screened by the $\tau$ factor, 138 failed to converged and are absent from the dataset. Beyond convergence failures,  the tolerance factor is an imperfect stability criterion. The choice of $\tau$ cutoff involves a balance of false positives and false negatives. The $\tau$-factor is linked to a stability probability using Platt scaling in Ref.~\citenum{bartelNewToleranceFactor2019}, and the cutoff of 4.18 corresponds to a 50\% stability probability, so some misclassified compositions near the boundary are to be expected. One illustration of this is provided by Kent \textit{et al.} who report the synthesis and analysis of Cs$_2$Na$Ln$I$_6$ with $Ln$=(Ce, Nd, Gd, Tb, Dy)\cite{kent_elusive_2023}, whereas the $\tau$-factor only predicts $Ln$=(Ce, Nd) to be stable. The remaining lanthanide iodide compositions exceed the maximum $\tau$-value by 0.02 or less, yet are evidently stable. 

\subsection{Structural Data}
We used the ICSD \cite{hellenbrandt_inorganic_2004} to compare the computed lattice parameters from our database with experimental values. We found 352 entries matching the HDP chemical formula, corresponding to 160 unique compositions. Filtering for entries with the $Fm\bar{3}m$ space group resulted in a total of 284 entries for 125 unique compositions. The results of the comparison are shown in Figure~\ref{fig:alat_spilling}a. Multiple ICSD entries for the same composition are shown with a vertical bar spanning the full range of values, and the markers indicate the average. Overall, there is good agreement between our simulations and experimental values. The mean absolute error (MAE) between the averaged ICSD and computed lattice parameters is 0.075\,{\AA}. The composition with the widest spread in lattice parameters is $\mathrm{Cs_2AgBiBr_6}$ with 24 entries measured at temperatures from 30\,K to 473\,K (Cs$_2$AgBiBr$_6$ features a structural phase transition from $Fm\bar{3}m$ to $I4/ m$ at 122\,K \cite{schadeStructuralOpticalProperties2019}). The 35 compositions without a cubic phase entry mostly have entries with the following space group symbols: $P\bar{3}m1$ (15 compositions), $R\bar{3}mH$ (12 compositions), and $I4/mmm$ (3 compositions). The trigonal crystal systems ($P\bar{3}m1$ and $R\bar{3}mH$ space groups) have octahedra with different connectivities. The remaining 5 crystal structures are differently distorted variants of the trigonal or cubic crystal systems with lower symmetries. While we cannot compare the $I4/mmm$ entries structurally to our simulations, we nevertheless analyze our bonding results within the $Fm\bar{3}m$ space group. Two of the $I4/mmm$ entries, $\mathrm{Cs_2CuHgCl_6}$ and $\mathrm{Cs_2PdHgCl_6}$, feature alternating elongated and contracted octahedra due to a Jahn-Teller distortion \cite{schroder_darstellung_1991,ji_jahnteller_2022}. Both also exhibit high directional asymmetry in our bonding analysis. Using the ICOBI-based directional asymmetry descriptor, we find: $\mathrm{Cs_2CuHgCl_6}$ with an asymmetry of 0.30 on copper and 0.12 on mercury, and $ \mathrm{Cs_2PdHgCl_6}$ with 0.05 on palladium and 0.13 on mercury. Using the ICOHPs as input for the bonding descriptor yields similar results. The third $I4/mmm$ entry ($\mathrm{Cs_2CuF_6}$) is missing from our database. These examples highlight the additional insights that can be provided by bonding analysis.

\subsection{Bonding Analysis}
\begin{figure}
    \centering
    \includegraphics[width=1\linewidth]{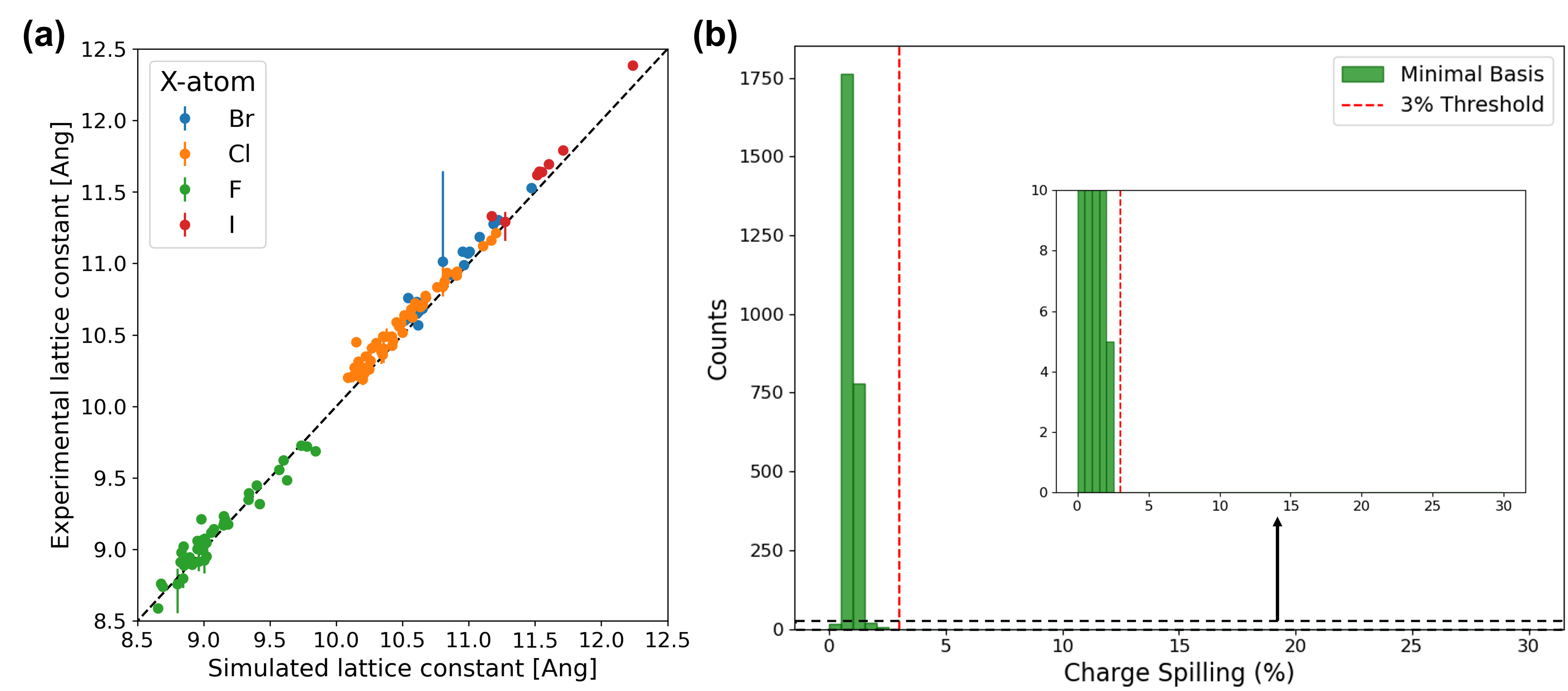}
    \caption{Validation of the data. (a) Comparison of experimental lattice constants from the ICSD with calculated lattice constants. Only entries with $Fm\bar{3}m$ space group are included (125/160 compositions). For compositions with multiple entries, the average is shown, with error bars indicating the full range. (b) Histogram of charge spillings in \textsc{lobster} projection, the most important quality measure. The inset shows a zoom into the black dashed box.}
    \label{fig:alat_spilling}
\end{figure}

Projecting the PAW wavefunction onto LCAO can lead to the loss or relocation of electrons. The extent of the relocation is measured by the absolute charge spilling\cite{maintz_lobster_2016}
\begin{equation}
    S_Q= \frac{1}{N_j} \sum_{\mathbf{k}}^{N_{\mathbf{k}}} w_{\mathbf{k}} \sum_{j}^{N_j}\left|1-O_{jj}    (\mathbf{k})\right|,
\end{equation}
where $O_{jj}$ are elements of the overlap matrix of the projected bands, which is summed over all occupied bands $j$, and $\mathbf k$-points with weight $w_{\mathbf{k}}$. The absolute charge spilling is the most important metric to assess the quality of the projection. Figure~\ref{fig:alat_spilling}b shows a histogram of the charge spillings in the dataset using the minimal basis set. None of the minimal basis projections exceed our imposed 3\% charge-spilling limit. Some compositions have high \textsc{bandOverlap} values (indicating deviations from the identity matrix during orthonormalization). High band overlap is of minor consequence to the projected values. Including additional basis functions, however, can directly affect ICOHP and ICOBI values, even if those orbitals are only fractionally occupied. Inspecting the results for the compositions with high band overlap values showed no notable irregularities. The results and analysis presented here are therefore based on the minimal basis set for all compositions. The projection quality measures are reported for all compositions (see HDP\_LobsterInfo.csv). Where additional basis-function calculations were performed, the projection data are also available on NOMAD (see \citenum{luc_walterbos_nomad_nodate}).

\subsection{Band gaps}
We also compare our calculated band-gap values with experimentally measured gaps. Experimental band gap determination can be method- and sample-dependent, leading to varying band gap values for materials with nominally identical composition and structure \cite{maughan_defect_2016}. We therefore provide only one or two sources per composition and focus on results obtained using the Tauc-plot method. The Tauc-plot method analyzes the UV-vis absorption onset and can give insight into whether a material has a direct or indirect band gap. This method has limitations, especially for (poly-)crystalline samples \cite{klein_limitations_2023}, but we use it here due to its prevalence in the literature. Note that we do not expect HSE06 band gaps to yield exact agreement with experiment. Our aim is instead to quantify the extent to which our calculated band gaps agree with experiment, within reasonable error bars given the limitations of our methods. Discrepancies between calculated and experimental values may be caused by limitations of the HSE06 functional - which uses a fixed amount of screened exact exchange - and from our neglect of dynamic distortions, electron-hole interactions, and spin-orbit coupling, which have all been shown to substantially affect band gap predictions of halide perovskites \cite{Scherpelz2016, Wiktor2017a, leppertPredictiveBandGaps2019, leppertExcitonsMetalhalidePerovskites2024, filipHalidePerovskitesFirst2024}. Given these caveats, our results for selected compositions, shown in Table~\ref{tab:gaps}, are in good agreement with experimental values and can serve as a starting point for future calculations using more accurate methods.  

\begin{table}[h!]
\begin{tabular}{|l|l|l|l|l|}
\hline
\cellcolor[HTML]{78E156}\textbf{Composition} & \cellcolor[HTML]{78E156}\textbf{Calc. $E_g$(eV)} & \cellcolor[HTML]{78E156}\textbf{Exp. $E_g$
(eV)} & \cellcolor[HTML]{78E156}\textbf{Notes} & \cellcolor[HTML]{78E156}\textbf{Ref.}\\ \hline
$\mathrm{Cs_2AgBiCl_6}$ & 2.75 & 2.50-2.92 (indirect)& single crystal & \citenum{choudhary_impact_2025},\citenum{bechir_lead-free_2023} \\ \hline
$\mathrm{Cs_2AgBiBr_6}$ & 2.07 & 2.02 (indirect) & single crystal & \citenum{choudhary_impact_2025} \\ \hline
$\mathrm{Cs_2AgInCl_6}$ & 2.32 & 3.53 (direct) & polycrystalline & \citenum{tran_designing_2017} \\ \hline
$\mathrm{Cs_2AgSbCl_6}$ & 2.13  & 2.54 (indirect) & polycrystalline & \citenum{tran_designing_2017} \\ \hline
$\mathrm{Cs_2NaBiCl_6}$ & 4.74 & 3.56 (indirect) & nanocrystals & \citenum{jiang_improved_2023} \\ \hline
$\mathrm{Cs_2NaBiBr_6}$ & 3.92 & 3.10 (indirect) & nanocrystals & \citenum{lamba_lead-free_2021} \\ \hline
$\mathrm{Cs_2CuBiCl_6}$ & 2.05 & 1.31 (indirect) & single crystal& \citenum{bechir_lead-free_2023} \\ \hline
$\mathrm{Cs_2CuSbCl_6}$ & 1.62 & 1.66 (indirect) & nanocrystals & \citenum{zhou_lead-free_2020} \\ \hline
$\mathrm{Cs_2SnI_6}$ & 0.77 & 1.25 (optical gap) & UV-vis reflectance & \citenum{maughan_defect_2016} \\ \hline
$\mathrm{Cs_2TeI_6}$ & 1.83 & 1.59 (optical gap) & UV-vis reflectance & \citenum{maughan_defect_2016} \\ \hline

\end{tabular}
\caption{Comparison of band gaps calculated with the HSE06 functional with experimental values.}
\label{tab:gaps}
\end{table}

\section{Data Availability}
The full calculation outputs can be accessed through NOMAD\cite{luc_walterbos_nomad_nodate}. The extracted data described in the Data Record section can be found on Zenodo\cite{lwalterbos_luccerboihdp_workflow_analysis_2026}. The UMAP explorer app, including all projections performed for this work is also deposited on Zenodo\cite{lwalterbos_luccerboihdp_umap_explorer_2026}.

\section{Code Availability}
All code developed for running the workflow, and analysing the results can be found on Zenodo together with the extracted data\cite{lwalterbos_luccerboihdp_workflow_analysis_2026}. The code used for the UMAP explorer app is contained in the same Zenodo Archive\cite{lwalterbos_luccerboihdp_umap_explorer_2026}. Both Zenodo archives also contain exact python package version information in the included requirements.txt files. \textsc{vasp} calculations were performed using version 6.3.0. \textsc{lobster} calculations with version 5.1.1.

\begin{acknowledgement}
L.W. and J.G. were supported by ERC Grant MultiBonds (grant agreement no. 101161771; funded by the European Union. Views and opinions expressed are however those of the author(s) only and do not necessarily reflect those of the European Union or the European Research Council Executive Agency. Neither the European Union nor the granting authority can be held responsible for them.) L.W. and J.G. gratefully thank the Gauss Centre for Supercomputing e.V. (www.gausscentre.eu) for providing generous computing time on the GCS Supercomputer SuperMUC-NG at the Leibniz Supercomputing Centre (www.lrz.de) under Project No. pn73da. L.L. acknowledges discussions with David Prendergast about the UMAP method and funding by the Dutch Research Council (NWO) through Grant No.~OCENW.M20.337 and VI.Vidi.223.072. We acknowledge the support of the NFDI consortium FAIRmat, funded by the Deutsche Forschungsgemeinschaft (DFG) – Project 460197019, for providing research data infrastructure and services. We thank Aakash A. Naik for adapting the NOMAD LOBSTER parsers to reflect recent LOBSTER updates and providing further technical support.

\end{acknowledgement}

\clearpage

\begin{longtable}{|l | l | p{0.55\linewidth}|}
\caption{Data contained in HDP\_BasicInfo.csv. \{site\} means the quantity is
given for A, B1, B2, and X separately. \{B1/B2\} means the quantity is reported
for B1 and B2 separately.}
\label{tab:basicinfo}\\

\hline
\cellcolor[HTML]{78E156}\textbf{Variable name} &
\cellcolor[HTML]{78E156}\textbf{Datatype} &
\cellcolor[HTML]{78E156}\textbf{Description} \\
\hline
CompID\_num              & int      & Unique ID number assigned to each composition, starting from 1000. \\ \hline
comp\_name\_simple       & string   & Chemical formula of composition without numbers (i.e., $\mathrm{CsBB'X}$). \\ \hline
comp\_name\_full         & string   & Full chemical formula of composition (i.e., $\mathrm{Cs_2BB'X_6}$). \\ \hline
element.\{site\}         & string   & Element symbol of \{site\}. \\ \hline
specie.\{site\}          & string   & Element and oxidation state of \{site\} (e.g., $\mathrm{Ag^{2+}}$). \\ \hline
r\_ionic.\{site\}        & float    & Shannon ionic radius in picometer associated with \{site\}. \\ \hline
tau\_factor              & float    & Bartel's $\tau$-factor \cite{bartelNewToleranceFactor2019} for this composition. \\ \hline
oct\_factor              & float    & Octahedral factor based on the average B-site ionic radius. \\ \hline
oct\_mismatch            & float    & Octahedral mismatch between the two B-sites introduced in \citenum{Filip2018}. \\ \hline
gen\_t\_factor           & float    & Generalised $t$-factor introduced in \citenum{Filip2018}. \\ \hline
geom\_stable             & bool     & Whether the compound is stable based on the geometric criteria of \citenum{Filip2018}. \\ \hline
broken\_condition        & string   & Indicates which geometric limit is broken if predicted non-stable by \citenum{Filip2018}. \\ \hline
block.\{B1/B2\}          & string   & Block of the periodic table \{B1/B2\} comes from (s/p/d/f). \\ \hline
row.\{B1/B2\}            & int      & Row of the periodic table \{B1/B2\} originates from. \\ \hline
inputcharge.\{B1/B2\}    & int      & Charge assigned to \{B1/B2\} in compound generation. \\ \hline
used\_lobbasis           & string   & \texttt{basis\{n\}} indicating which set of atomic orbitals was used. \\ \hline
used\_lobbasis\_funcs    & dict(str)& Dictionary of atomic orbitals used for each element. \\ \hline
\end{longtable}

\begin{longtable}{| l | l | p{0.55\linewidth}|}
\caption{Data contained in HDP\_StructuralInfo.csv. \{B1/B2\} means the quantity
is given for B1 and B2 separately.}
\label{tab:structuralinfo}\\

\hline
\cellcolor[HTML]{78E156}\textbf{Variable name} &
\cellcolor[HTML]{78E156}\textbf{Datatype} &
\cellcolor[HTML]{78E156}\textbf{Description} \\ \hline
lattice\_a\_primitive      & float & Lattice constant of the primitive unit cell in \AA. \\ \hline
lattice\_a\_conventional   & float & Lattice constant of the conventional unit cell in \AA. \\ \hline
distance\_\{B1/B2\}\_X    & float & Distance between \{B1/B2\} and X ions in \AA. \\ \hline
distance\_\{B1/B2\}\_A    & float & Distance between \{B1/B2\} and A cations in \AA. \\ \hline
size\_Oh\_\{B1/B2\}       & float & Size of the \{B1/B2\} octahedron, measured from opposing X anions in \AA. \\ \hline

\end{longtable}

\begin{longtable}{| l | p{0.12\linewidth} | l | p{0.42\linewidth}|}
\caption{Data contained in bandedgeInfo\_lsodos.csv. Data is collected for
spin-up (``1'') and spin-down (``-1'') channels separately, and combined
(``comb'') using the highest VBM and lowest CBM across channels. \{site\} means
the quantity is included for A, B1, B2, and X separately.}
\label{tab:bandedgeinfo} \\ \hline
\cellcolor[HTML]{78E156}\textbf{Variable name} &
\cellcolor[HTML]{78E156}\textbf{Spin} &
\cellcolor[HTML]{78E156}\textbf{Datatype} &
\cellcolor[HTML]{78E156}\textbf{Description} \\ \hline
\endfirsthead

\multicolumn{4}{l}{\tablename\ \thetable{} -- \textit{continued}} \\[0.5em]
\hline
\cellcolor[HTML]{78E156}\textbf{Variable name} &
\cellcolor[HTML]{78E156}\textbf{Spin} &
\cellcolor[HTML]{78E156}\textbf{Datatype} &
\cellcolor[HTML]{78E156}\textbf{Description} \\
\hline
\endhead

\hline
\multicolumn{4}{r}{\textit{Continued on next page}} \\
\endfoot

\hline
\endlastfoot

VBM                    & ``1'', ``-1'', ``comb'' & float  & Valence band maximum energy (eV). \\ \hline
CBM                    & ``1'', ``-1'', ``comb'' & float  & Conduction band minimum energy (eV). \\ \hline
band\_gap              & ``1'', ``-1'', ``comb'' & float  & Band gap energy (eV). \\ \hline
VBM.totcontr.\{site\}  & ``1'', ``-1'', ``comb'' & float  & Total contribution of \{site\} to the valence band. \\ \hline
VBM.orbcontr.\{site\}  & ``1'', ``-1'', ``comb'' & float  & Highest single orbital contribution of \{site\} to the valence band. \\ \hline
VBM.orbital.\{site\}   & ``1'', ``-1'', ``comb'' & string & Main contributing orbital of \{site\} to the valence band. \\ \hline
VBM.orbchar.\{site\}   & ``1'', ``-1'', ``comb'' & string & Character (s/p/d/f) of the main contributing orbital of \{site\} to the valence band. \\ \hline
VBM.orborder.\{site\}  & ``1'', ``-1'', ``comb'' & string & Principal quantum number ($n$) of the main contributing orbital of \{site\} to the valence band. \\ \hline
CBM.totcontr.\{site\}  & ``1'', ``-1'', ``comb'' & float  & Relative contribution of \{site\} to the conduction band. \\ \hline
CBM.orbcontr.\{site\}  & ``1'', ``-1'', ``comb'' & float  & Highest single orbital contribution of \{site\} to the conduction band. \\ \hline
CBM.orbital.\{site\}   & ``1'', ``-1'', ``comb'' & string & Main contributing orbital of \{site\} to the conduction band. \\ \hline
CBM.orbchar.\{site\}   & ``1'', ``-1'', ``comb'' & string & Character (s/p/d/f) of the main contributing orbital of \{site\} to the conduction band. \\ \hline
CBM.orborder.\{site\}  & ``1'', ``-1'', ``comb'' & string & Principal quantum number ($n$) of the main contributing orbital of \{site\} to the conduction band. \\ \hline
spin\_cbm              & ``combined''            & string & Spin channel of the overall CBM. \\ \hline
spin\_vbm              & ``combined''            & string & Spin channel of the overall VBM. \\ \hline
spin\_forbidden        & ``combined''            & bool   & Whether the VBM and CBM are in differing spin channels, with a minimal energy difference of 50\,meV. \\ \hline
cond\_type             & ``1'', ``-1'', ``comb'' & string & Whether the spin channel is metallic ($E_g \leq 100$\,meV), semiconducting, insulating ($E_g > 4$\,eV), or half-metallic (combined channel only). \\ \hline

\end{longtable}

\begin{longtable}{|p{0.28\linewidth}| l | p{0.52\linewidth}|}
\caption{Data contained in LobsterInfo.csv. \{B1/B2\} means the
quantity is reported for B1 and B2 separately. \{x/y/z\} indicates that the
quantity is determined separately along the Cartesian axes.}
\label{tab:lobsterinfo}\\

\hline
\cellcolor[HTML]{78E156}\textbf{Variable name} &
\cellcolor[HTML]{78E156}\textbf{Datatype} &
\cellcolor[HTML]{78E156}\textbf{Description} \\
\hline
\endfirsthead

\multicolumn{3}{l}{\tablename\ \thetable{} -- \textit{continued}} \\[0.5em]
\hline
\cellcolor[HTML]{78E156}\textbf{Variable name} &
\cellcolor[HTML]{78E156}\textbf{Datatype} &
\cellcolor[HTML]{78E156}\textbf{Description} \\
\hline
\endhead

\hline
\multicolumn{3}{r}{\textit{Continued on next page}} \\
\endfoot

\hline
\endlastfoot

abs\_charge\_spilling         & float & The absolute charge spilling from the \textsc{lobster} projection (\%). \\ \hline
bandOverlap\_exists            & bool  & Whether \textsc{lobster} wrote the bandOverlaps file, indicating imperfect orthonormalization. \\ \hline
bandOverlap\_maxDev            & float & Maximum deviation from identity matrix recorded in bandOverlaps. \\ \hline
bandOverlap\_perc\_kpts        & float & Percentage of k-points that exceed a deviation of 0.1 from the identity matrix (\%). \\ \hline
popdiff.total                  & float & Total difference between spin-up and spin-down gross populations. \\ \hline
popdiff.B1                     & float & Spin-up\&down population difference on B1-site. \\ \hline
popdiff.B2                     & float & Spin-up\&down population difference on B2-site. \\ \hline
charge.\{site\}                & float & Net charge on \{site\}. \\ \hline
charge.total                   & float & Sum of local charges. \\ \hline
charge.balanced                & bool  & Whether the total structure is charge neutral with a tolerance of $0.1e^{-}$. \\ \hline
Icohp.\{B1/B2\}.sum            & float & Sum of the 6 B-X ICOHP values on site \{B1/B2\} (eV). \\ \hline
Icohp.\{B1/B2\}.avg            & float & Average of the 6 B-X ICOHP values on site \{B1/B2\} (eV). \\ \hline
Icohp.\{B1/B2\}. avg\_spinup    & float & Average of the 6 B-X ICOHP values on site \{B1/B2\}, based on spin-up electrons only (eV). \\ \hline
Icohp.\{B1/B2\}. avg\_spindown  & float & Average of the 6 B-X ICOHP values on site \{B1/B2\}, based on spin-down electrons only (eV). \\ \hline
Icohp.\{B1/B2\}. \{x/y/z\}\_avg      & float & Average of the B-X ICOHP values aligned with \{x/y/z\}-cartesian axis (eV). \\ \hline
Icohp.\{B1/B2\}. \{x/y/z\}\_spinup   & float & Average of the B-X spin-up ICOHP values aligned with \{x/y/z\}-cartesian axis (eV). \\ \hline
Icohp.\{B1/B2\}. \{x/y/z\}\_spindown & float & Average of the B-X spin-down ICOHP values aligned with \{x/y/z\}-cartesian axis (eV). \\ \hline
Icobi.\{B1/B2\}.sum            & float & Sum of the 6 B-X ICOBI values on site \{B1/B2\}. \\ \hline
Icobi.\{B1/B2\}.avg            & float & Average of the 6 B-X ICOBI values on site \{B1/B2\}. \\ \hline
Icobi.\{B1/B2\}. avg\_spinup    & float & Average of the 6 B-X ICOBI values on site \{B1/B2\}, based on spin-up electrons only. \\ \hline
Icobi.\{B1/B2\}. avg\_spindown  & float & Average of the 6 B-X ICOBI values on site \{B1/B2\}, based on spin-down electrons only. \\ \hline
Icobi.\{B1/B2\}. \{x/y/z\}\_avg      & float & Average of the B-X ICOBI values aligned with \{x/y/z\}-cartesian axis. \\ \hline
Icobi.\{B1/B2\}. \{x/y/z\}\_spinup   & float & Average of the B-X spin-up ICOBI values aligned with \{x/y/z\}-cartesian axis. \\ \hline
Icobi.\{B1/B2\}. \{x/y/z\}\_spindown & float & Average of the B-X spin-down ICOBI values aligned with \{x/y/z\}-cartesian axis. \\ \hline
\{Icohp/Icobi\}.axial\_asym \_index.\{B1/B2\} & float & Norm of absolute differences in bonding indicator along each axis (from \citenum{belli_chemical_2025}).\\ \hline
\{Icohp/Icobi\}.directional \_asym\_index.\{B1/B2\} & float & Relative differences in bonding indicator between cartesian axes.\\ \hline
MadelungEnergy                 & float & Total electrostatic stabilization energy (eV). \\ \hline
Sitepotential.\{B1/B2\}        & float & Electrostatic stabilization coming from site \{B1/B2\} (eV). \\ \hline
\end{longtable}

\begingroup
\small
\begin{longtable}{llllll}
\caption{Overview of PAW pseudopotentials used with corresponding electronic
configuration, and atomic orbital basis. The first two columns indicate whether
relaxation was performed with a reduced-valence PAW (RPAW) that included additional frozen
electrons. Boldfaced PAW names indicate cases where the chosen PAW deviates
from that recommended by \textsc{vasp}.}
\label{tab:paw} \\

\toprule
\textbf{Element} & \textbf{RPAW} & \textbf{config.} & \textbf{PAW}
& \textbf{config.} & \textbf{LCAO basis} \\
\midrule
\endfirsthead

\multicolumn{6}{l}{\tablename\ \thetable{} -- \textit{continued}} \\[0.5em]
\toprule
\textbf{Element} & \textbf{RPAW} & \textbf{config.} & \textbf{PAW}
& \textbf{config.} & \textbf{LCAO basis} \\
\midrule
\endhead

\midrule
\multicolumn{6}{r}{\textit{Continued on next page}} \\
\endfoot

\bottomrule
\endlastfoot

Li  & & & Li\_sv              & $1s^{2} 2s^{1}$                                    & $1s$ $2s$               \\
Be  & & & Be                  & $2s^{1.99} 2p^{0.01}$                              & $2s$                    \\
B   & & & B                   & $2s^{2} 2p^{1}$                                    & $2p$ $2s$               \\
N   & & & N                   & $2s^{2} 2p^{3}$                                    & $2p$ $2s$               \\
F   & & & F                   & $2s^{2} 2p^{5}$                                    & $2p$ $2s$               \\
Na  & & & Na\_pv              & $2p^{6} 3s^{1}$                                    & $2p$ $3s$               \\
Mg  & & & Mg                  & $3s^{2}$                                           & $3s$                    \\
Al  & & & Al                  & $3s^{2} 3p^{1}$                                    & $3p$ $3s$               \\
Si  & & & Si                  & $3s^{2} 3p^{2}$                                    & $3p$ $3s$               \\
P   & & & P                   & $3s^{2} 3p^{3}$                                    & $3p$ $3s$               \\
Cl  & & & Cl                  & $3s^{2} 3p^{5}$                                    & $3p$ $3s$               \\
K   & & & K\_sv               & $3s^{2} 3p^{6} 4s^{1}$                            & $3p$ $3s$ $4s$          \\
Ca  & & & Ca\_sv              & $3s^{2} 3p^{6} 4s^{2}$                            & $3p$ $3s$ $4s$          \\
Sc  & & & Sc\_sv              & $3s^{2} 3p^{6} 3d^{2} 4s^{1}$                    & $3d$ $3p$ $3s$ $4s$     \\
Ti  & & & Ti\_sv              & $3s^{2} 3p^{6} 3d^{3} 4s^{1}$                    & $3d$ $3p$ $3s$ $4s$     \\
V   & & & V\_sv               & $3s^{2} 3p^{6} 3d^{4} 4s^{1}$                    & $3d$ $3p$ $3s$ $4s$     \\
Cr  & & & Cr\_pv              & $3p^{6} 3d^{5} 4s^{1}$                            & $3d$ $3p$ $4s$          \\
Mn  & & & Mn\_pv              & $3p^{6} 3d^{6} 4s^{1}$                            & $3d$ $3p$ $4s$          \\
Fe  & & & Fe                  & $3d^{7} 4s^{1}$                                   & $3d$ $4s$               \\
Co  & & & Co                  & $3d^{8} 4s^{1}$                                   & $3d$ $4s$               \\
Ni  & & & Ni                  & $3d^{9} 4s^{1}$                                   & $3d$ $4s$               \\
Cu  & & & Cu                  & $3d^{10} 4s^{1}$                                  & $3d$ $4s$               \\
Zn  & & & Zn                  & $3d^{10} 4s^{2}$                                  & $3d$ $4s$               \\
Ga  & & & Ga\_d               & $3d^{10} 4s^{2} 4p^{1}$                           & $3d$ $4p$ $4s$          \\
Ge  & & & Ge\_d               & $3d^{10} 4s^{2} 4p^{2}$                           & $3d$ $4p$ $4s$          \\
As  & & & As                  & $4s^{2} 4p^{3}$                                   & $4p$ $4s$               \\
Se  & & & Se                  & $4s^{2} 4p^{4}$                                   & $4p$ $4s$               \\
Br  & & & Br                  & $4s^{2} 4p^{5}$                                   & $4p$ $4s$               \\
Rb  & & & Rb\_sv              & $4s^{2} 4p^{6} 4d^{0.001} 5s^{0.999}$            & $4p$ $4s$               \\
Sr  & & & Sr\_sv              & $4s^{2} 4p^{6} 4d^{0.001} 5s^{1.999}$            & $4p$ $4s$ $5s$          \\
Y   & & & Y\_sv               & $4s^{2} 4p^{6} 4d^{2} 5s^{1}$                    & $4d$ $4p$ $4s$ $5s$     \\
Zr  & & & Zr\_sv              & $4s^{2} 4p^{6} 4d^{3} 5s^{1}$                    & $4d$ $4p$ $4s$ $5s$     \\
Nb  & & & Nb\_sv              & $4s^{2} 4p^{6} 4d^{4} 5s^{1}$                    & $4d$ $4p$ $4s$ $5s$     \\
Mo  & & & Mo\_sv              & $4s^{2} 4p^{6} 4d^{5} 5s^{1}$                    & $4d$ $4p$ $4s$ $5s$     \\
Tc  & & & Tc\_pv              & $4p^{6} 4d^{6} 5s^{1}$                            & $4d$ $4p$ $5s$          \\
Ru  & & & Ru\_pv              & $4p^{6} 4d^{7} 5s^{1}$                            & $4d$ $4p$ $5s$          \\
Rh  & & & Rh\_pv              & $4p^{6} 4d^{8} 5s^{1}$                            & $4d$ $4p$ $5s$          \\
Pd  & & & Pd                  & $4d^{9} 5s^{1}$                                   & $4d$ $5s$               \\
Ag  & & & Ag                  & $4d^{10} 5s^{1}$                                  & $4d$ $5s$               \\
Cd  & & & Cd                  & $4d^{10} 5s^{2}$                                  & $4d$ $5s$               \\
In  & & & In\_d               & $4d^{10} 5s^{2} 5p^{1}$                           & $4d$ $5p$ $5s$          \\
Sn  & & & Sn\_d               & $4d^{10} 5s^{2} 5p^{2}$                           & $4d$ $5p$ $5s$          \\
Sb  & & & Sb                  & $5s^{2} 5p^{3}$                                   & $5p$ $5s$               \\
Te  & & & Te                  & $5s^{2} 5p^{4}$                                   & $5p$ $5s$               \\
I   & & & I                   & $5s^{2} 5p^{5}$                                   & $5p$ $5s$               \\
Cs  & & & \textbf{Cs\_sv\_GW} & $5s^{2} 5p^{6} 5d^{1}$                            & $5p$ $5s$               \\
Ba  & & & Ba\_sv              & $5s^{2} 5p^{6} 5d^{0.01} 6s^{1.99}$             & $5p$ $5s$ $6s$          \\
La  & & & La                  & $4f^{0.0001} 5s^{2} 5p^{6} 5d^{0.9999} 6s^{2}$  & $5p$ $5s$ $6s$          \\
Ce  & \textbf{Ce\_3} & $5s^{2} 5p^{6} 5d^{1} 6s^{2}$ & Ce   & $4f^{1} 5s^{2} 5p^{6} 5d^{1} 6s^{2}$             & $4f$ $5d$ $5p$ $5s$ $6s$\\
Pr  & Pr\_3 & $5s^{2} 5p^{6} 5d^{1} 6s^{2}$ & \textbf{Pr}   & $4f^{2.5} 5s^{2} 5p^{6} 5d^{0.5} 6s^{2}$        & $4f$ $5p$ $5s$ $6s$     \\
Nd  & Nd\_3 & $5s^{2} 5p^{6} 5d^{1} 6s^{2}$ & \textbf{Nd}   & $4f^{3.5} 5s^{2} 5p^{6} 5d^{0.5} 6s^{2}$        & $4f$ $5p$ $5s$ $6s$     \\
Pm  & Pm\_3 & $5s^{2} 5p^{6} 5d^{1} 6s^{2}$ & \textbf{Pm}   & $4f^{4.5} 5s^{2} 5p^{6} 5d^{0.5} 6s^{2}$        & $4f$ $5p$ $5s$ $6s$     \\
Sm  & Sm\_3 & $5s^{2} 5p^{6} 5d^{1} 6s^{2}$ & \textbf{Sm}   & $4f^{5.5} 5s^{2} 5p^{6} 5d^{0.5} 6s^{2}$        & $4f$ $5p$ $5s$ $6s$     \\
Eu  & Eu\_3 & $5p^{6} 5d^{1} 6s^{2}$         & \textbf{Eu}   & $4f^{6.5} 5s^{2} 5p^{6} 5d^{0.5} 6s^{2}$        & $4f$ $5p$ $5s$ $6s$     \\
Gd  & Gd\_3 & $5p^{6} 5d^{1} 6s^{2}$         & \textbf{Gd}   & $4f^{7.5} 5s^{2} 5p^{6} 5d^{0.5} 6s^{2}$        & $4f$ $5p$ $5s$ $6s$     \\
Tb  & Tb\_3 & $5p^{6} 5d^{1} 6s^{2}$         & \textbf{Tb}   & $4f^{8.5} 5s^{2} 5p^{6} 5d^{0.5} 6s^{2}$        & $4f$ $5p$ $5s$ $6s$     \\
Dy  & Dy\_3 & $5p^{6} 5d^{1} 6s^{2}$         & \textbf{Dy}   & $4f^{9.5} 5s^{2} 5p^{6} 5d^{0.5} 6s^{2}$        & $4f$ $5p$ $5s$ $6s$     \\
Ho  & Ho\_3 & $5p^{6} 5d^{1} 6s^{2}$         & \textbf{Ho}   & $4f^{10.5} 5s^{2} 5p^{6} 5d^{0.5} 6s^{2}$       & $4f$ $5p$ $5s$ $6s$     \\
Er  & Er\_3 & $5p^{6} 5d^{1} 6s^{2}$         & \textbf{Er}   & $4f^{11.5} 5s^{2} 5p^{6} 5d^{0.5} 6s^{2}$       & $4f$ $5p$ $5s$ $6s$     \\
Tm  & Tm\_3 & $5p^{6} 5d^{1} 6s^{2}$         & \textbf{Tm}   & $4f^{12.5} 5s^{2} 5p^{6} 5d^{0.5} 6s^{2}$       & $4f$ $5p$ $5s$ $6s$     \\
Yb  & Yb\_3 & $5p^{6} 5d^{1} 6s^{2}$         & \textbf{Yb}   & $4f^{13.5} 5s^{2} 5p^{6} 5d^{0.5} 6s^{2}$       & $4f$ $5p$ $5s$ $6s$     \\
Lu  & Lu\_3 & $5p^{6} 5d^{1} 6s^{2}$         & \textbf{Lu}   & $4f^{14} 5s^{2} 5p^{6} 5d^{1} 6s^{2}$            & $4f$ $5d$ $5p$ $5s$ $6s$\\
Hf  & Hf\_pv & $5p^{6} 5d^{3} 6s^{1}$        & \textbf{Hf\_sv} & $5s^{2} 5p^{6} 5d^{4}$                         & $5d$ $5p$ $5s$          \\
Ta  & & & Ta\_pv              & $5p^{6} 5d^{4} 6s^{1}$                            & $5d$ $5p$ $6s$          \\
W   & & & W\_sv               & $5s^{2} 5p^{6} 5d^{5} 6s^{1}$                    & $5d$ $5p$ $5s$ $6s$     \\
Re  & & & Re                  & $5d^{6} 6s^{1}$                                   & $5d$ $6s$               \\
Os  & & & Os                  & $5d^{7} 6s^{1}$                                   & $5d$ $6s$               \\
Ir  & & & Ir                  & $5d^{8} 6s^{1}$                                   & $5d$ $6s$               \\
Pt  & & & Pt                  & $5d^{9} 6s^{1}$                                   & $5d$ $6s$               \\
Au  & & & Au                  & $5d^{10} 6s^{1}$                                  & $5d$ $6s$               \\
Hg  & & & Hg                  & $5d^{10} 6s^{2}$                                  & $5d$ $6s$               \\
Tl  & & & Tl\_d               & $5d^{10} 6s^{2} 6p^{1}$                           & $5d$ $6p$ $6s$          \\
Pb  & & & Pb\_d               & $5d^{10} 6s^{2} 6p^{2}$                           & $5d$ $6p$ $6s$          \\
Bi  & & & Bi\_d               & $5d^{10} 6s^{2} 6p^{3}$                           & $5d$ $6p$ $6s$          \\
Po  & & & Po\_d               & $5d^{10} 6s^{2} 6p^{4}$                           & $5d$ $6p$ $6s$          \\
At  & & & At                  & $6s^{2} 6p^{5}$                                   & $6p$ $6s$               \\
Rn  & & & Rn                  & $6s^{2} 6p^{6}$                                   & $6p$ $6s$               \\
Fr  & & & Fr\_sv              & $6s^{2} 6p^{6} 7s^{1}$                            & $6p$ $6s$ $7s$          \\
Ra  & & & Ra\_sv              & $6s^{2} 6p^{6} 7s^{2}$                            & $6p$ $6s$ $7s$          \\
Ac  & & & Ac                  & $6s^{2} 6p^{6} 6d^{1} 7s^{2}$                    & $6d$ $6p$ $6s$ $7s$     \\
Th  & & & Th                  & $5f^{1} 6s^{2} 6p^{6} 6d^{1} 7s^{2}$             & $5f$ $6d$ $6p$ $6s$ $7s$\\
Pa  & & & Pa                  & $5f^{1} 6s^{2} 6p^{6} 6d^{2} 7s^{2}$             & $5f$ $6d$ $6p$ $6s$ $7s$\\
U   & & & U                   & $5f^{2} 6s^{2} 6p^{6} 6d^{2} 7s^{2}$             & $5f$ $6d$ $6p$ $6s$ $7s$\\
Np  & & & Np                  & $5f^{3} 6s^{2} 6p^{6} 6d^{2} 7s^{2}$             & $5f$ $6d$ $6p$ $6s$ $7s$\\
Pu  & & & Pu                  & $5f^{4} 6s^{2} 6p^{6} 6d^{2} 7s^{2}$             & $5f$ $6d$ $6p$ $6s$ $7s$\\
Am  & & & Am                  & $5f^{5} 6s^{2} 6p^{6} 6d^{2} 7s^{2}$             & $5f$ $6d$ $6p$ $6s$ $7s$\\
Cm  & & & Cm                  & $5f^{6} 6s^{2} 6p^{6} 6d^{2} 7s^{2}$             & $5f$ $6d$ $6p$ $6s$ $7s$\\
\end{longtable}
\endgroup

\clearpage

\providecommand{\latin}[1]{#1}
\makeatletter
\providecommand{\doi}
  {\begingroup\let\do\@makeother\dospecials
  \catcode`\{=1 \catcode`\}=2 \doi@aux}
\providecommand{\doi@aux}[1]{\endgroup\texttt{#1}}
\makeatother
\providecommand*\mcitethebibliography{\thebibliography}
\csname @ifundefined\endcsname{endmcitethebibliography}  {\let\endmcitethebibliography\endthebibliography}{}

\end{document}